\documentclass[sigconf]{acmart}

\usepackage{longfbox}
\usepackage{graphicx}
\usepackage{subfigure}
\usepackage{amsmath}
\usepackage[american]{babel}
\usepackage{subcaption}
\usepackage[most]{tcolorbox}
\usepackage{balance} 
\usepackage{enumitem} 
\usepackage{color}
\usepackage{url}
\usepackage{bm}
\usepackage{mathrsfs}
\usepackage{multirow}
\usepackage{colortbl}
\usepackage{algorithm}
\usepackage{algpseudocode}
\usepackage{threeparttable} 
\usepackage{graphicx}
\usepackage{caption}
\makeatletter \let\c@lofdepth\relax \makeatother
\makeatletter \let\c@lotdepth\relax \makeatother
\usepackage{microtype}
\usepackage{graphicx}
\usepackage{tabularray}
\usepackage{xspace}
\usepackage{makecell}
\usepackage{multirow}
\usepackage{algorithm}
\usepackage{algpseudocode}
\usepackage{wrapfig,booktabs}
\usepackage{amssymb}
\usepackage{listings}
\usepackage{fancyhdr}
\definecolor{lightgreen}{rgb}{0.56, 0.93, 0.56} 
\definecolor{lightcoral}{rgb}{1.0, 0.5, 0.31}  

\newcommand{\system}{\textsc{LaTCoder}\xspace}%
\newcommand{\systemnospace}{\textsc{LaTCoder}}
\newcommand{\benchmark}{CC-HARD\xspace}%

\newcommand{\mypara}[1]{\smallskip \noindent\textbf{#1} \xspace}

\AtBeginDocument{%
  }

\copyrightyear{2025}
\acmYear{2025}
\setcopyright{acmlicensed}\acmConference[KDD '25]{Proceedings of the 31st ACM
SIGKDD Conference on Knowledge Discovery and Data Mining V.2}{August 3--7,
2025}{Toronto, ON, Canada}
\acmBooktitle{Proceedings of the 31st ACM SIGKDD Conference on Knowledge
Discovery and Data Mining V.2 (KDD '25), August 3--7, 2025, Toronto, ON, Canada}
\acmDOI{10.1145/3711896.3737016}
\acmISBN{979-8-4007-1454-2/2025/08}

\author{Yi Gui}
\authornote{Also with National Engineering Research Center for Big Data Technology and System, Services Computing Technology and System Lab, Cluster and Grid Computing Lab, School of Computer Science and Technology, Huazhong University of Science and Technology, Wuhan, 430074, China.}
\authornote{These authors contribute equally to this research.}
\orcid{0009-0006-2841-7942}
\affiliation{%
  \institution{Huazhong University of Science and Technology}
  \city{Wuhan}
  \country{China}
}

\author{Zhen Li}
\authornotemark[1]
\authornotemark[2]

\orcid{0009-0007-0873-6126}
\affiliation{%
   \institution{Huazhong University of Science and Technology}
  \city{Wuhan}
  \country{China}
}

\author{Zhongyi Zhang}
\authornotemark[2]
\orcid{0009-0009-9951-3335}
\affiliation{%
  \institution{Huazhong University of Science and Technology}
  \city{Wuhan}
  \country{China}
}

\author{Guohao Wang}
\authornotemark[2]
\orcid{0009-0004-1046-1510}
\affiliation{%
  \institution{Huazhong University of Science and Technology}
  \city{Wuhan}
  \country{China}
}

\author{Tianpeng Lv}
\orcid{0009-0006-2860-4211}
\affiliation{%
  \institution{Huazhong University of Science and Technology}
  \city{Wuhan}
  \country{China}
}

\author{Gaoyang Jiang}
\orcid{0009-0009-1032-3659}
\affiliation{%
  \institution{Huazhong University of Science and Technology}
  \city{Wuhan}
  \country{China}
}

\author{Yi Liu}
\orcid{0009-0002-7120-3158}
\affiliation{%
  \institution{Huazhong University of Science and Technology}
  \city{Wuhan}
  \country{China}
}

\author{Dongping Chen}
\orcid{0009-0009-9848-2557}
\affiliation{%
  \institution{Huazhong University of Science and Technology}
  \city{Wuhan}
  \country{China}
}

\author{Yao Wan}
\authornotemark[1]
\authornote{Yao Wan is the corresponding author (\texttt{wanyao@hust.edu.cn}).}

\orcid{0000-0001-6937-4180}
\affiliation{%
   \institution{Huazhong University of Science and Technology}
  \city{Wuhan}
  \country{China}
}

\author{Hongyu Zhang}
\orcid{0000-0002-3063-9425}
\affiliation{%
  \institution{Chongqing University}
  \city{Chongqing}
  \country{China}
}

\author{Wenbin Jiang}
\authornotemark[1]
\orcid{0000-0001-5628-8806}
\affiliation{%
   \institution{Huazhong University of Science and Technology}
  \city{Wuhan}
  \country{China}
}

\author{Xuanhua Shi}
\authornotemark[1]
\orcid{0000-0001-8451-8656}
\affiliation{%
   \institution{Huazhong University of Science and Technology}
  \city{Wuhan}
  \country{China}
}

\author{Hai Jin}
\authornotemark[1]
\orcid{0000-0002-3934-7605}
\affiliation{%
   \institution{Huazhong University of Science and Technology}
  \city{Wuhan}
  \country{China}
}

\title{\system: Converting Webpage Design to Code with Layout-as-Thought}

\renewcommand{\shortauthors}{Yi Gui et al.}
\begin{document}

\begin{abstract}
Converting webpage designs into code (design-to-code) plays a vital role in \textit{User Interface} (UI) development for front-end developers, bridging the gap between visual design and functional implementation.
While recent \textit{Multimodal Large Language Models} (MLLMs) have shown significant potential in design-to-code tasks, they often fail to accurately preserve the layout during code generation.
To this end, we draw inspiration from the \textit{Chain-of-Thought} (CoT) reasoning in human cognition and propose \system, a novel approach that enhances layout preservation in webpage design during code generation with \textit{Layout-as-Thought} (\textsc{LaT}).
Specifically, we first introduce a simple yet efficient algorithm to divide the webpage design into image blocks. Next, we prompt MLLMs using a CoT-based approach to generate code for each block. Finally, we apply two assembly strategies—absolute positioning and an MLLM-based method—followed by dynamic selection to determine the optimal output.
We evaluate the effectiveness of \system using multiple backbone MLLMs (i.e., DeepSeek-VL2, Gemini, and GPT-4o) on both a public benchmark and a newly introduced, more challenging benchmark (\benchmark) that features complex layouts.
The experimental results on automatic metrics demonstrate significant improvements. Specifically, TreeBLEU scores increased by 66.67\% and MAE decreased by 38\% when using DeepSeek-VL2, compared to direct prompting. 
Moreover, the human preference evaluation results indicate that annotators favor the webpages generated by \system in over 60\% of cases, providing strong evidence of the effectiveness of our method.
\end{abstract}



\maketitle

\newcommand\kddavailabilityurl{https://doi.org/10.5281/zenodo.15515215}


\section{Introduction}
Front-end developers typically write webpage code based on \textit{Graphical User Interface} (GUI) mockups created by UI designers. This process involves translating visual components—such as elements, layouts, and functionalities—into \textit{Hypertext Markup Language} (HTML), \textit{Cascading Style Sheets} (CSS), and JavaScript code, which is often time-consuming and costly.
Due to the significant burden of generating large amounts of repetitive webpage code, more than 75.8\% of front-end developers have adopted AI tools to improve development efficiency\footnote{\url{https://tsh.io/state-of-frontend/}}. Consequently, there is an increasing need for automated design-to-code solutions that can transform a webpage design into code. 

Previously, several efforts aimed to generate UI code from simple-styled design images (such as hand-drawn sketches~\cite{Alex2019_Sketch2code}) using smaller models~\cite{Tony2018_pix2code}. Recently, the powerful capabilities of \textit{Multimodal Large Language Models} (MLLMs), such as Pix2Struct~\cite{Kenton2023_Pix2Struct}, GPT-4V~\cite{Ouyang2022TrainingLM}, and Claude~\cite{TheC3}, have made it possible to directly convert high-resolution webpage designs into code.
Several works have curated large-scale corpora for training purposes, including WebSight~\cite{Laurenccon2024UnlockingTC}, WebCode2M~\cite{gui2024vision2ui}, and Web2Code~\cite{DBLP:journals/corr/abs-2406-20098}.
Notably, \citet{Si2024Design2CodeHF} established a high-quality benchmark and introduced a novel metric (i.e., visual score) specifically tailored for evaluating the performance of MLLMs. Building on these foundational efforts, subsequent studies have focused on fine-tuning task-specific MLLMs~\cite{Laurenccon2024UnlockingTC, gui2024vision2ui} or instructing interactive MLLMs with augmented methods such as self-revision~\cite{Si2024Design2CodeHF}.
In addition, \citet{wanAutomaticallyGeneratingUI2024} proposed DCGen, which improves code generation by synthesizing code directly from the entire design while incorporating natural language descriptions for subregions.

\begin{figure}[!t]
    \centering
    \includegraphics[width=0.95\linewidth]{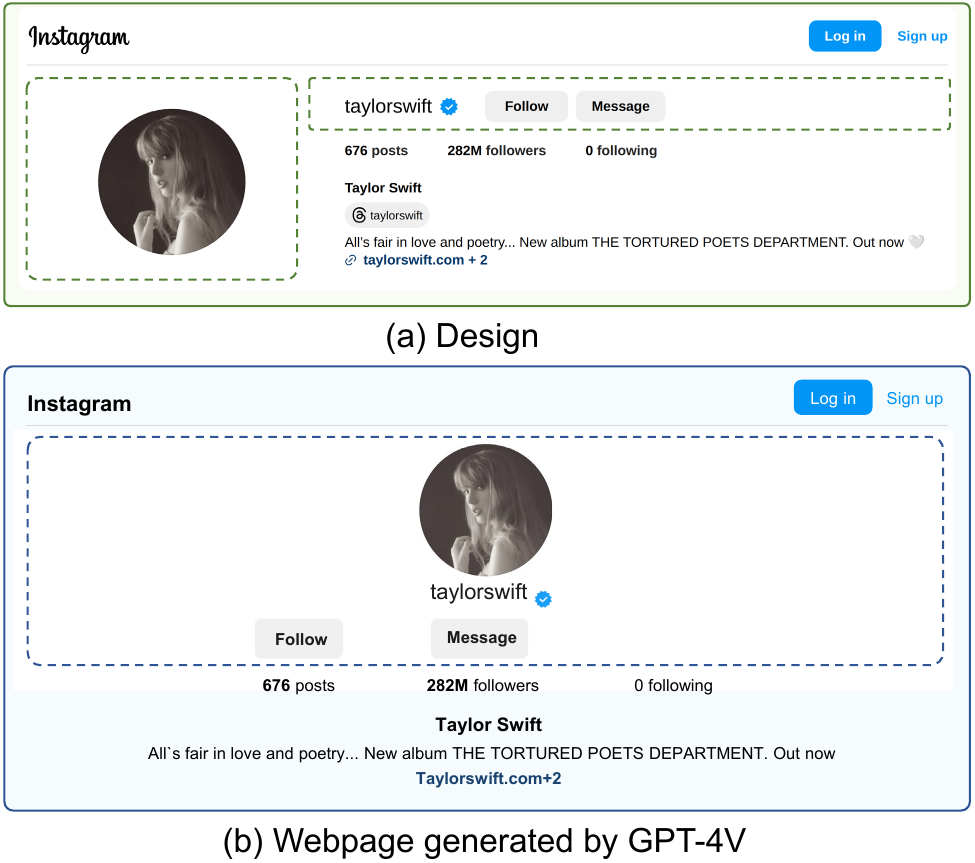}
    \caption{
    A real-world bad case from the famous project, \textit{screenshot-to-code}~\protect\cite{homepage_screen2shot}, where GPT-4V incorrectly arranges the elements during generation (as highlighted in boxes).
    }
    \vspace{-1em}
    \label{fig_motivation}
\end{figure}

Despite the promising performance achieved by previous studies, we are still far from fully automating UI synthesis for real-world webpages. Existing methods primarily rely on \textit{monolithic} generation, where the complete webpage code is generated directly from the design using MLLMs. However, we observe that this approach is inherently limited, as \textit{partial layout information is often lost during code generation}. Consequently, MLLMs struggle to preserve the original structure and layout when dealing with real-world webpages that contain diverse styles and complex content.

\begin{figure*}[!t]
    \centering
    \includegraphics[width=0.98\linewidth]{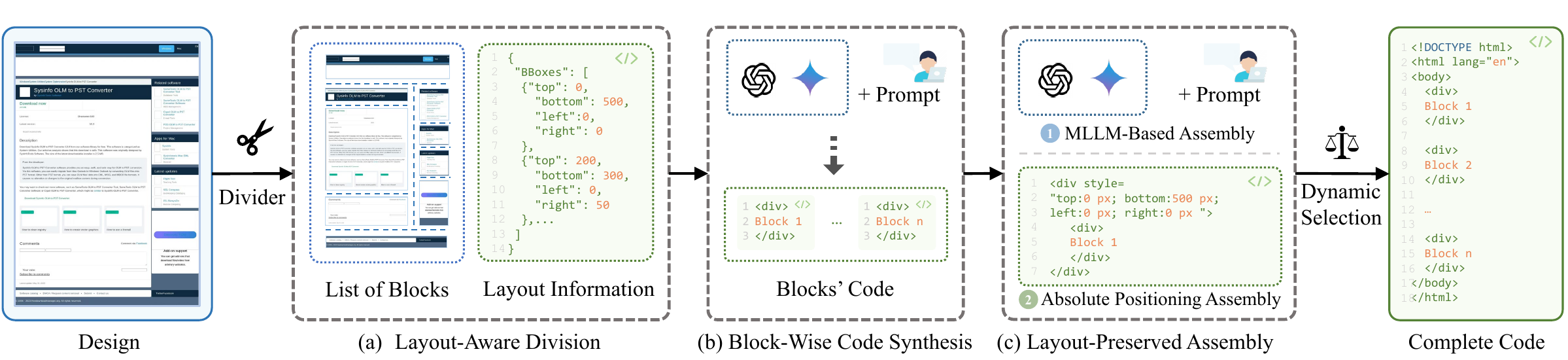}
    \vspace{-1em}
    \caption{The workflow of \system.}
    \vspace{-1em}
    \label{fig_overview}
\end{figure*}

For better illustration, Figure~\ref{fig_motivation} presents a real-world example (with minor modifications) from the well-known project \textit{screen-to-shot}~\cite{homepage_screen2shot}, where MLLMs generate corresponding code from website screenshots using meticulously crafted prompts.
Figure~\ref{fig_motivation}(a) displays a screenshot from an Instagram page, while Figure~\ref{fig_motivation}(b) shows the corresponding webpage synthesized by GPT-4V.
As observed, the layout of the synthesized webpage differs significantly from the original design, particularly in the elements highlighted in boxes. 
GPT-4V incorrectly arranges the elements vertically instead of horizontally, even when instructed to ``\textit{Make sure to always get the layout right}''.  
Our further investigation reveals that this limitation of easily losing partial layout information when translating designs into webpage code is not an exception but a limitation shared by many MLLMs.
We speculate that this limitation may be due to the vulnerabilities~\cite{Bai2024HallucinationOM} of MLLMs in factual interpretation~\cite{DBLP:conf/fat/BenderGMS21} and numerical reasoning~\cite{DBLP:conf/nips/HendrycksBKABTS21}, which affects accurate capture of element positions and sizes in webpage designs.
 
\mypara{Our Work.}
To mitigate the limitation of MLLMs in accurately capturing layout information, we propose a novel approach called \system.
Our method incorporates an efficient divider, a CoT-based code generator, and a flexible code assembler.
Drawing inspiration from the success of \textit{Chain-of-Thought} (CoT), which decomposes complex tasks into simpler steps solvable by LLMs in sequence, we introduce a similar concept: \textit{Layout-as-Thought} (\textsc{LaT}), which generates webpage code block by block, as opposed to traditional monolithic generation.
In this approach, the design is broken down into a series of image blocks, each regarded as a ``thought'' that can be processed independently.
We begin by dividing the webpage design into distinct image blocks in subregions, represented as bounding boxes (BBoxes). Using these BBoxes, the design is cropped into image blocks, which are then input into MLLMs one at a time for subregion code generation via CoT-based prompts. Next, guided by the BBoxes, we assemble the generated code for each image block using both absolute positioning and MLLM-based assembly to form the complete webpage code.
Finally, we introduce a verifier to validate the results from both assembly strategies, further enhancing performance.
Since each block is anchored by its BBox within the overall layout, \system significantly alleviates the limitation of MLLMs in capturing layout information accurately.
Furthermore, by breaking the monolithic generation process into block-by-block code generation, \system reduces the burden on MLLMs to generate lengthy code, resulting in improved accuracy in detail generation.
   
To validate the effectiveness and generalizability of \systemnospace, we introduce a more challenging dataset, featuring more complex layouts. 
Specifically, we manually sample from the Common Crawl dataset~\cite{ccdataset} and generate paired data to curate the dataset, which we refer to as \benchmark.
We evaluate our method and four state-of-the-art baseline approaches using different backbone MLLMs on a public dataset (i.e., Design2Code-Hard) and our newly introduced \benchmark. 
Experimental results show that integrating various MLLMs as backbones into our method significantly boosts performance across all automatic metrics, particularly on TreeBLEU~\cite{gui2024vision2ui}, which measures the similarity of sub-structures in the HTML \textit{Document Object Model} (DOM) tree.
Moreover, we conduct a pairwise human preference evaluation of our method against each baseline, using majority voting from six annotators. The results of this evaluation demonstrate that, in most cases, human annotators prefer the webpages generated by our method, providing strong evidence for its effectiveness.

\mypara{Contributions.}
The key contributions of this paper are as follows:
\begin{itemize}[leftmargin=*, itemsep=0.05mm]
    \item We propose \system, a novel approach that enhances layout preservation in converting webpage designs to code using MLLMs with \textsc{LaT}.
    \item To assess the effectiveness of MLLMs, we introduce a new benchmark named \benchmark\footnote{\url{https://huggingface.co/datasets/xcodemind/CC-HARD}}, featuring more complex layouts, driving further advancements in MLLMs for webpage design-to-code generation.
    \item We conduct extensive experiments to validate the effectiveness of our approach, evaluated on the Design2Code-HARD and \benchmark datasets. All the materials are available at: \url{https://github.com/CGCL-codes/naturalcc/tree/main/examples/latcoder}. 
\end{itemize}

\section{Design-to-Code: The Problem}
The design-to-code task aims to translate design images—such as screenshots of existing websites or design mockups created by designers—into corresponding code. A typical modern webpage consists of three core components: HTML, which defines structural elements (e.g., \texttt{<div>}, \texttt{<button>}, \texttt{<img>}) and their hierarchical relationships; CSS, which controls layout properties (e.g., position, flexbox, grid) and visual styling (e.g., margin, padding, font size) to determine spatial arrangement and rendering; and JavaScript, which implements functionalities such as event handling and dynamic content updates.
Our work focuses on static webpage synthesis from designs, where the rendered appearance is jointly determined by the HTML structure and CSS styling. Consequently, webpages with the same HTML code can have entirely different appearances depending on their CSS.
The key of layout preservation in the design-to-code task with MLLMs is to accurately capture and map the size and position information of elements in the design into HTML/CSS code.

Given a high-resolution webpage design, \system aims to automatically generate the corresponding HTML and CSS code. The resulting webpage, once rendered, should closely resemble the input design in terms of layout, styling, and content. 
The design-to-code task can be defined as a mapping function \( F: I \rightarrow (H, S) \), where
$I\in \mathbb{R}^{M \times N \times 3}$ is the input webpage design with a size of $M \times N$,
$H = \{ h_1, h_2, \dots, h_n \} \quad$represents the generated HTML elements, and $S = \{ s_j \mid s_j = (p_j, v_j) \} \quad$ specifies the CSS rules. Here, $(p_j, v_j)$ denotes a pair of property and value in CSS.
The objective is to minimize the visual difference between the rendered output $R = Render(H \cup S) $ and the input $I$:
$$
\hat{H}, \hat{S} = \arg \min_{H, S} \, D(I, R)\,,
$$
where $\hat{H}$, $\hat{S}$ are the optimal HTML and CSS that minimize the perceptual distance $D(I, R)$ between the input design $I$ and the rendered output $R$.
 
\section{\system: Our Approach}
\label{sec_division}
Figure~\ref{fig_overview} illustrates the workflow of our proposed \system, which is composed of three components: (a) layout-aware division, (b) block-wise code synthesis, and (c) layout-preserved assembly. 
We first divide the design into smaller image blocks using an algorithm that ensures text integrity while recording the corresponding BBox information.
Furthermore, we instruct MLLMs to generate code for each block using CoT-based prompts.
Finally, we assemble the generated block code using both absolute positioning and MLLM-based strategies, followed by dynamic selection to get the best output.
By anchoring blocks to their original positions in the design, this approach maximizes layout preservation while significantly reducing the burden on MLLMs when generating lengthy code.
\begin{table}[!h]
\caption{Statistics of non-standard cases across three public datasets: IL refers to irregular layouts with misaligned or unevenly spaced elements; OL indicates overlapping layouts; GB stands for gradient backgrounds.}
\begin{tabular}{lccccc}
\toprule[1pt]
Dataset          & Size & IL & OL & GB & Total \\
\midrule
Design2Code      & 485  & 11        & 5          & 5         & 21 (4.33\%)              \\
Design2Code-HARD & 80   & 5         & 1          & 5         & 11 (13.75\%)             \\
WebCode2M-Long   & 256  & 1         & 6          & 1         & 8 (3.13\%)              \\
\bottomrule[1pt]
\end{tabular}
\label{fig_non_standard}
\vspace{-1em}
\end{table}

\subsection{Layout-Aware Division}
\begin{figure}[!t]
    \centering
    \includegraphics[width=0.99\linewidth]{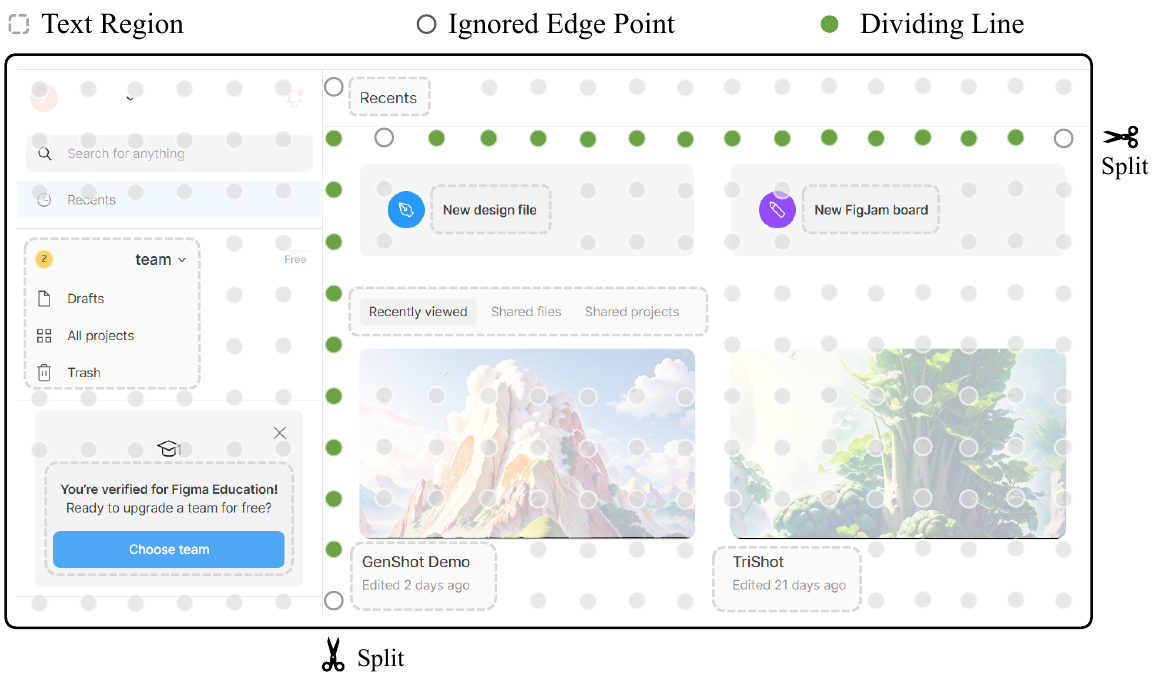}
    \caption{A toy example of dividing line detection.}
    \vspace{-1em}
    \label{fig_split}
\end{figure}

Dividing the webpage design into appropriately sized image blocks is an essential step for generating code incrementally.
Traditional image segmentation methods like Mask R-CNN~\cite{DBLP:journals/pAMI/HeGDG20}  and U-Net~\cite{DBLP:conf/miccai/RonnebergerFB15} detect irregular object boundaries through instance or semantic segmentation~\cite{Shelhamer2014FullyCN}.
In contrast, the vast majority of webpage layouts follow CSS box model conventions (as illustrated in Table~\ref{fig_non_standard}), where structured, rectangular containers form the fundamental building blocks.
Therefore, we propose a specialized and efficient algorithm for detecting horizontal/vertical lines to divide design mockups into grid-aligned blocks, better aligning with the structured nature of HTML/CSS layout systems.

\mypara{Dividing Line Detection.}
To divide the design image into appropriately sized and grid-aligned blocks, we aim to identify a set of horizontal/vertical lines that can fully partition the image.
There are other ways~\cite{wanAutomaticallyGeneratingUI2024} to define and search for such dividing lines, but our focus is on exploring the potential of \textsc{LaT} in webpage generation, rather than finding the optimal dividing algorithm.
Therefore, we adopt a simple yet effective definition for the dividing lines: horizontal or vertical solid-colored lines, with the distance to the nearest adjacent dividing line no greater than a predefined threshold $\tau$. To achieve this, we design a search algorithm (as shown in Algorithm~\ref{alg_image_segmentation}), which scans line by line in only one direction (either horizontal or vertical) to find the set of dividing lines. The algorithm is applied recursively to the original design image and each sub-image until no further dividing lines can be found, ultimately gathering all dividing lines.
\begin{algorithm}[!h]
\caption{Get dividing lines in image $I$}
\begin{algorithmic}[1]
    \Require Image $I$, minimum distance threshold $\tau$
    \Ensure A set $S$ of dividing lines
    \State Initialize $h \gets \text{left edge}$, $v \gets \text{top edge}$
    \State Initialize empty set $S$
    
    \For{each row starting from $h$}
        \State Let $h1$ be the next candidate line
        
        \If{\texttt{IsSolidColored}$(h1)$ \textbf{and} \texttt{Distance}$(h, h1)$ $\geq \tau$}
            \State Add $h1$ to $S$
            \State Update $h \gets h1$
        \EndIf
    \EndFor
    
    \If{$S \neq \emptyset$}
        \State \textbf{Return} $S$
    \EndIf
    
    \For{each column starting from $v$}
        \State Let $v1$ be the next candidate line
        \If{\texttt{IsSolidColored}$(v1)$ \textbf{and} $\texttt{Distance}(v, v1)$ $\geq \tau$}
            \State Add $v1$ to $S$
            \State Update $v \gets v1$
        \EndIf
    \EndFor
    
    \State \textbf{Return} $S$
\end{algorithmic}
\label{alg_image_segmentation}
\end{algorithm}

\mypara{Algorithm Optimizations.}
In practice, we propose the following optimizations to enhance the accuracy and efficiency of the detection algorithm, as exemplified in Figure~\ref{fig_split}.
\textbf{(1) Ensuring text region integrity.}
Dividing lines may split text regions, particularly when they span across line or paragraph gaps. To avoid this, we integrate \textit{ Optical Character Recognition} (OCR) to detect text regions and ensure that a line is only considered a valid dividing line if it does not go through any text regions.
\textbf{(2) Improving the efficiency via grid sampling.}
Pixel-level scanning on high-resolution design images is inefficient and computationally expensive. Instead, we apply a grid sampling technique, where pixels on the image are sampled at fixed intervals when determining a solid-colored line, rather than scanning at the step of every individual pixel.
\textbf{(3) Ignoring a few points at edges.}
Pixels in the borders of images may hinder the determination of a solid-colored line. To mitigate this, we ignore the first few pixels at the edges during the determination.

With all dividing lines detected, the design is divided into a list of blocks of subregions, represented as BBoxes. Blocks that are smaller than a predefined threshold $\theta$ will be merged into adjacent blocks. These BBoxes are recorded further for block-wise code generation and assembly. 
More details can be found in Section~\ref{sec_divider_details}.

\smallskip
\noindent\textbf{\textsc{\underline{A Toy Example.}}}
Figure~\ref{fig_split} displays a toy example of detecting dividing lines.
First, we sample points at fixed intervals across the entire image, forming a grid.
Next, we scan the grid row by row or column by column to identify potential dividing lines.
During this process, a valid dividing line must be a solid-colored line that does not cross any text regions.
To minimize the impact of borders, we ignore one edge pixel in the determination.
As a result, the algorithm detects one vertical and one horizontal segmentation line, marked by green points in the figure.
Notably, although the edge pixel is ignored, it is still considered part of the dividing line.

\subsection{Block-Wise Code Synthesis}
\begin{figure}[!t]
    \centering
    \includegraphics[width=0.99\linewidth]{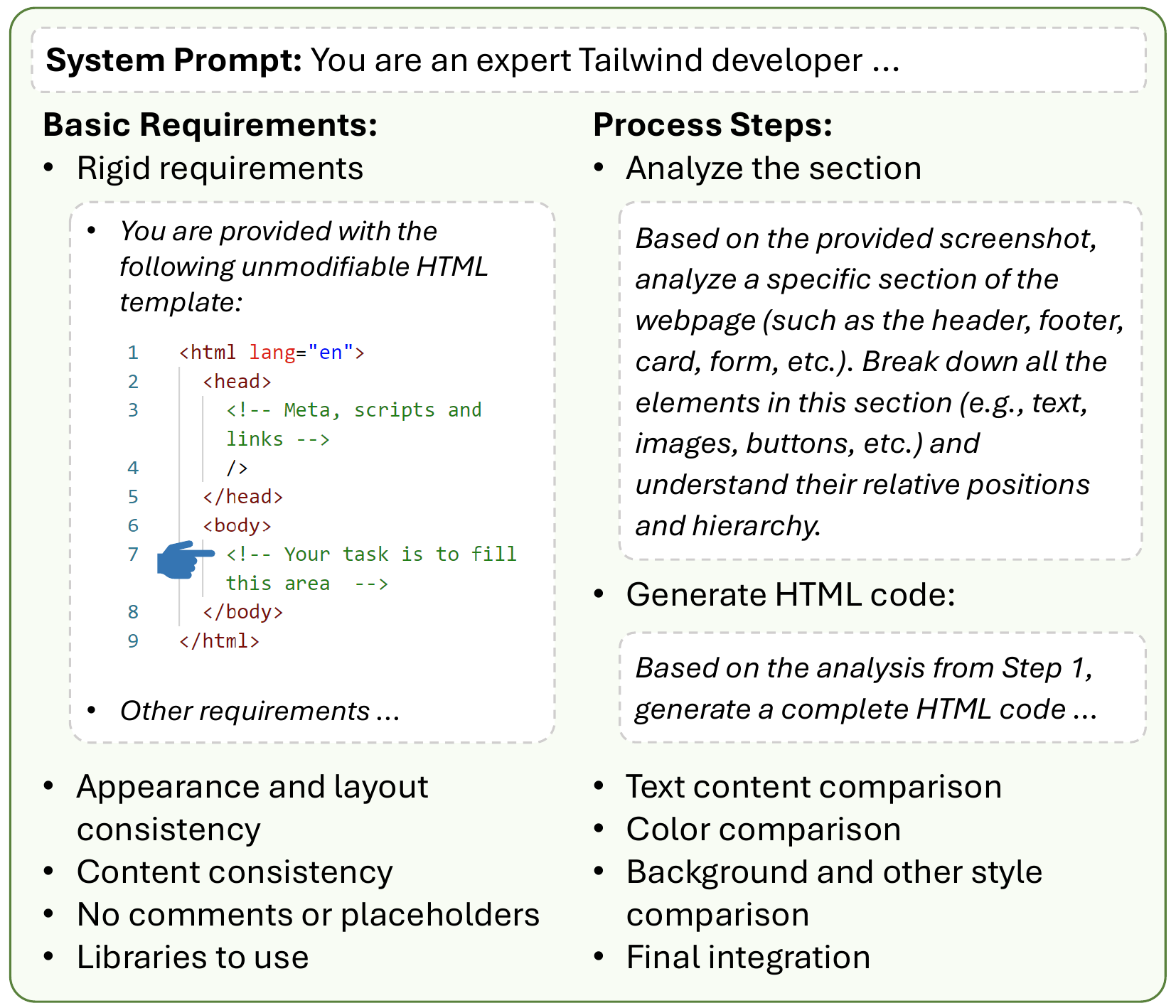}
    \caption{The simplified prompt for generating image block code (the full version is shown in Figure~\ref{fig_prompt_generate} in Appendix).}
    \vspace{-1em}
    \label{fig_prompt_sketch}
\end{figure}

The goal of this module is to generate code snippets for image blocks. 
We first crop out all image blocks from the design using their BBoxes, and then input each image block into interactive  MLLMs one by one for code generation.
We have meticulously designed a prompt for generating accurate and high-quality webpage code. Our prompt (see Figure~\ref{fig_prompt_generate} in Appendix) design mainly adheres to the following principles:
\textbf{(1) Use a unified webpage template.} Generate Tailwind-style HTML/CSS code for a \texttt{div} within a fixed webpage template. This unified template ensures that the styles of all blocks' code remain consistent after assembling.
\textbf{(2) Layout first.} Focus first on the appearance and layout consistency, then strive to maintain content consistency.
\textbf{(3) CoT-based generation.} Generate step by step: analyze the image block, generate initial HTML/CSS code, check the code on text content, color, background, and other styles, then polish the code to finalize the generation.
For weaker models, such as DeepSeek-VL2, due to the shorter context limit, we provide a slightly simplified version of the prompt, which can be found in the artifacts. 
For better illustration, we provide a simplified version of the complete prompt in Figure~\ref{fig_prompt_sketch}.

\subsection{Layout-Preserved Assembly}
We explore two distinct strategies for assembling the code of all image blocks into complete code while preserving the overall design layout: absolute positioning and MLLM-based assembly.
MLLM-based assembly is more flexible but requires that MLLMs support a longer context to handle the code of all image blocks effectively.
In contrast, absolute positioning assembly is faster and particularly suitable for weaker MLLMs with shorter context windows.
Each strategy offers unique advantages—absolute positioning excels in position accuracy, while MLLM-based assembly often produces more aesthetically pleasing results.
As a result, we retain both strategies where applicable and introduce dynamic selection to get the best output.

\mypara{Strategy 1: Absolute Positioning Assembly (APS).}
We utilize the coordinates in the BBoxes obtained during the division stage to perform absolute positioning assembly. Each image block's code is encapsulated within a parent \texttt{<div>}, with its position and size set according to the BBox.

\mypara{Strategy 2: MLLM-Based Assembly (MS).}
For interactive MLLMs, constraints can be applied by adjusting the prompt to generate more accurate and high-quality webpage code. Therefore, we design the prompt (as shown in the artifacts) following the main principles outlined below to guide the code assembly:
\textbf{(1) Preserve the layout:}
merge the provided blocks' code based on the original design image and the BBox information of each image block, ensuring that the positions match those in the design.
\textbf{(2) CoT-based assembly:}
analyze the layout information using all BBoxes, assemble the code, compare the generated webpage with the design image to ensure content, layout, and style consistency—fixing discrepancies if needed—then refine and finalize the complete code.

\mypara{Dynamic Strategy Selection.}
The dynamic strategy selection aims to determine the best output when both assembly strategies are available.
If the MLLM has a sufficiently long context window to process all image block codes, both strategies can be considered. However, when using a weaker MLLM, such as DeepSeek-VL2, only absolute positioning is employed due to its limited context capacity.
We design a verifier to evaluate the similarity between the generated webpage and the webpage design. Importantly, this evaluation is \textbf{reference-free}, meaning it does not rely on the source code of the target. Consequently, we compare the screenshots of the generated webpage and the original design, providing a practical evaluation mechanism.  
Inspired by~\cite{DBLP:journals/corr/abs-2406-20098}, we explore the potential of employing the \textit{MLLM-as-a-Judge} paradigm for implementing the verifier. 
Although \textit{MLLM-as-a-Judge} has achieved notable success across multiple domains~\cite{Chen2024MLLMasaJudgeAM, Dinh2024SciExBL, pu2025judge, chen2024interleaved}, we find it is not capable of reliably and accurately verifying the similarity or quality of images in our scenario, even with meticulously designed prompts (see them in artifacts).  
As a result, we return to traditional automatic metrics for assessing image similarity.  
Previous studies~\cite{wanMRWebExplorationGenerating2024} suggest that Mean Absolute Error (\textbf{MAE}) and Normalized Earth Mover’s Distance (\textbf{NEMD})~\cite{Rubner2000TheEM} align more closely with human preference on the webpage similarity. However, since both metrics operate at the pixel level, they may overlook semantic (high-level) information. To mitigate this limitation, we combine \textbf{MAE} and \textbf{CLIP}~~\cite{DBLP:conf/icml/RadfordKHRGASAM21} similarity into a composite metric for the verifier. The metric, called the verify score, is defined as follows:
\[
\text{Verify Score} = \frac{1}{2} \times \left(1 - \frac{\text{MAE}}{255}\right) + \frac{1}{2} \times \text{CLIP}\,.
\]  
We empirically set equal coefficients (0.5) for the two components in this formula.

\section{Experimental Setup}
\subsection{Evaluation Datasets}
\mypara{Design2Code-HARD.}
One of the primary benchmarks for webpage generation is Design2Code~\cite{Si2024Design2CodeHF}. However, according to the latest version of its paper, when using stronger models, such as GPT-4o, the performance gap between direct generation and various augmented methods on the Design2Code dataset has become minimal. As a result, we have adopted its newly proposed, more complex version—Design2Code-HARD—as one of our text benchmarks. This version includes 80 extremely long samples.

\begin{table}[!t]
\centering 
\renewcommand{\arraystretch}{1.1}
\caption{A statistical comparison between Design2Code-HARD and \benchmark.
}
\vspace{-1em}
\begin{tabular}{lcc}
\toprule[1pt]
                    & Design2Code-HARD                        & \benchmark   \\ 
\midrule
Size                & 80                                  & 128               \\
Avg. Len (tokens) & 8900±2399                         & 8416±2190         \\
Avg. Text Len (tokens) & 3554±2820                    & 969±762         \\
Avg. Tags      & 251±232                                & 274±66            \\
Avg. DOM Depth      & 10±4                                    & 16±3              \\
Avg. Unique Tags     & 23±5                            & 27±5              \\
\bottomrule[1pt]
\end{tabular}
\label{tb_dataset_stat}
\vspace{-1em}
\end{table}

\mypara{\benchmark: A More Challenging Benchmark.}
However, during testing, we found that the complexity of Design2Code-HARD mainly lies in the text length, while its layout and structure remain relatively simple. As a result, \textbf{the performance differences between various methods are still minimal on Design2Code-HARD}.
To address this, we instruct two experts to manually obtain more challenging samples from the Common Crawl~\cite{ccdataset} dataset and generate paired data, curating a new benchmark, called \benchmark.
We conduct a statistical analysis of the two benchmarks, as shown in Table~\ref{tb_dataset_stat}. While the overall average length distributions of the two datasets are similar, the textual content in Design2Code-HARD is much longer than in \benchmark. This suggests that the total length of \benchmark is consumed more by source code rather than textual content. Compared to Design2Code-HARD, \benchmark contains significantly more tags and unique tags, as well as a deeper DOM tree. These differences indicate that \benchmark is more challenging in terms of layout and structure, which is further supported by our experimental results.

\begin{table*}[!t]
\centering
\caption{
The overall performance of \system and baseline models with different backbone MLLMs across two datasets.
}

\vspace{-1em}
\begin{threeparttable}
\setlength{\tabcolsep}{2pt} 
\renewcommand{\arraystretch}{1.1}
\scalebox{1}{
\begin{tabular}{lcccccccc}
\toprule[1pt]
\multicolumn{1}{l|}{\multirow{2}{*}{Method}} & \multicolumn{4}{c|}{\textbf{Design2Code-HARD}}                                   & \multicolumn{4}{c}{\textbf{\benchmark}}                       \\
\multicolumn{1}{l|}{}                        & TreeBLEU($\uparrow$)  & CLIP($\uparrow$)      & Visual Score($\uparrow$) & \multicolumn{1}{c|}{MAE($\downarrow$)}         & TreeBLEU($\uparrow$)  & CLIP($\uparrow$)      & Visual Score($\uparrow$) & MAE($\downarrow$)         \\
\midrule
\multicolumn{9}{c}{\textit{\textbf{DeepSeek-VL2}}}                                                                                                                                           \\
\midrule
\multicolumn{1}{l|}{Direct Prompting}        & 0.12(±0.07) & \textbf{0.81(±0.08)} & 0.64(±0.25)    & \multicolumn{1}{c|}{69.13(±28.53)} & 0.09(±0.04) & \textbf{0.75(±0.10)} & 0.64(±0.30)    & 66.91(±21.30) \\
\multicolumn{1}{l|}{\textbf{\system} (APS)}                 & \textbf{0.19(±0.09)} & 0.77(±0.09) & \textbf{0.72(±0.19)}    & \multicolumn{1}{c|}{\textbf{51.63(±36.89)}} & \textbf{0.15(±0.05)} & 0.74(±0.11) & \textbf{0.72(±0.22)}    & \textbf{41.13(±24.41)} \\
\multicolumn{1}{l|}{$\Delta$}                & \colorbox {lightgreen}{58.33\% }       & \colorbox{lightcoral}{-4.94\%}        & \colorbox{lightgreen}{12.5\%}           & \multicolumn{1}{c|}{\colorbox{lightgreen}{-25.31\%}}          & \colorbox{lightgreen}{66.67\%}        & \colorbox{lightcoral}{-1.33\%}        & \colorbox{lightgreen}{12.5\%}           & \colorbox{lightgreen}{-38.53\%} \\
\midrule
\multicolumn{9}{c}{\textbf{\textit{Gemini}}}                                                                                                                                               \\
\midrule
\multicolumn{1}{l|}{Direct Prompting}        & 0.16(±0.09) & 0.84(±0.08) & 0.76(±0.19)    & \multicolumn{1}{c|}{65.52(±31.45)} & 0.09(±0.04) & 0.78(±0.10) & 0.76(±0.23)    & 65.22(±24.68) \\
\multicolumn{1}{l|}{Text-Augmented}          & 0.14(±0.09) & 0.83(±0.09) & 0.76(±0.17)    & \multicolumn{1}{c|}{63.34(±30.52)} & 0.09(±0.05) & 0.78(±0.09) & 0.74(±0.24)    & 66.02(±24.00) \\
\multicolumn{1}{l|}{Self-Revision}           & 0.15(±0.09) & 0.84(±0.08) & 0.77(±0.17)    & \multicolumn{1}{c|}{62.59(±30.50)} & 0.10(±0.05) & 0.77(±0.10) & 0.75(±0.23)    & 67.20(±23.70) \\
\multicolumn{1}{l|}{DCGen}                   & 0.14(±0.09) & 0.79(±0.09) & 0.74(±0.17)    & \multicolumn{1}{c|}{78.45(±29.94)} & 0.09(±0.05) &0.73(±0.12)  & 0.74(±0.24)    &	75.79(±25.79) \\
\multicolumn{1}{l|}{\textbf{\system} (MS)}         & \textbf{0.16(±0.07)} & 0.83(±0.08) & \textbf{0.81(±0.07)}    & \multicolumn{1}{c|}{59.72(±26.21)} & 0.13(±0.05) & 0.78(±0.10) & 0.76(±0.25)    & 58.52(±21.66) \\
\multicolumn{1}{l|}{\textbf{\system} (APS)}      & 0.16(±0.08) & 0.86(±0.07) & 0.78(±0.17)    & \multicolumn{1}{c|}{40.21(±25.02)} & 0.13(±0.05) & 0.80(±0.09) & 0.76(±0.25)    & 37.59(±18.85) \\
\multicolumn{1}{l|}{\textbf{\system} }          & 0.16(±0.08) & \textbf{0.86(±0.06)} & 0.78(±0.17)    & \multicolumn{1}{c|}{\textbf{37.50(±21.50)}} & \textbf{0.13(±0.04)} & \textbf{0.80(±0.09) }& \textbf{0.78(±0.23) }   & \textbf{37.15(±17.52)} \\
\multicolumn{1}{l|}{$\Delta$}                & \colorbox {lightgreen}{0\%}        & \colorbox {lightgreen}{2.38\%}       & \colorbox {lightgreen}{5.19\%}           & \multicolumn{1}{c|}{\colorbox {lightgreen}{-40.09\%}}          & \colorbox {lightgreen}{30\%}        & \colorbox {lightgreen}{2.56\%}        & \colorbox {lightgreen}{2.63\%}          & \colorbox {lightgreen}{-43.03\%} \\
\midrule
\multicolumn{9}{c}{\textbf{\textit{GPT-4o}}}                                                                                                                                                  \\
\midrule
\multicolumn{1}{l|}{Direct Prompting}        & 0.16(±0.11) & 0.84(±0.08) & 0.75(±0.19)    & \multicolumn{1}{c|}{61.62(±25.06)} & 0.09(±0.05) & 0.79(±0.10) & 0.76(±0.24)    & 66.18(±21.94) \\
\multicolumn{1}{l|}{Text-Augmented}          & 0.16(±0.11) & 0.86(±0.07) & 0.79(±0.17)    & \multicolumn{1}{c|}{54.21(±22.94)} & 0.10(±0.05) & 0.79(±0.11) & 0.78(±0.23)    & 64.88(±21.09) \\
\multicolumn{1}{l|}{Self-Revision}           & 0.16(±0.10) & 0.86(±0.07) & 0.79(±0.17)    & \multicolumn{1}{c|}{54.53(±24.23)} & 0.10(±0.05) & 0.79(±0.10) & 0.78(±0.23)    & 64.82(±21.11) \\
\multicolumn{1}{l|}{DCGen}                   & 0.17(±0.10) & 0.83(±0.10) & 0.77(±0.16)    & \multicolumn{1}{c|}{63.31(±27.21)} & 0.10(±0.05) & 0.74(±0.12) & 0.75(±0.26)    & 68.31(±21.58) \\
\multicolumn{1}{l|}{\textbf{\system} (MS)}         & 0.20(±0.11) & 0.86(±0.07) & \textbf{0.82(±0.12)}    & \multicolumn{1}{c|}{59.55(±25.84)} & 0.16(±0.06) & 0.78(±0.09) & 0.78(±0.25)    & 58.29(±20.45) \\
\multicolumn{1}{l|}{\textbf{\system} (APS)}      & 0.20(±0.11) & 0.86(±0.08) & 0.80(±0.15)    & \multicolumn{1}{c|}{36.21(±22.36)} & 0.16(±0.06) & 0.80(±0.09) & 0.80(±0.23)    & 37.53(±20.66) \\
\multicolumn{1}{l|}{\textbf{\system} }          & \textbf{0.20(±0.11)} & \textbf{0.87(±0.07)} & 0.80(±0.15)    & \multicolumn{1}{c|}{\textbf{33.93(±18.15)}} & \textbf{0.16(±0.06)} & \textbf{0.81(±0.09)} & \textbf{0.80(±0.23)}    & \textbf{36.80(±17.49)} \\
\multicolumn{1}{l|}{$\Delta$}                &  \colorbox {lightgreen}{17.65\%}        & \colorbox {lightgreen}{1.27\%}        & \colorbox {lightgreen}{3.8\%}           & \multicolumn{1}{c|}{\colorbox {lightgreen}{-37.41\%}}          & \colorbox {lightgreen}{60\%}        & \colorbox {lightgreen}{2.53\%}        & \colorbox {lightgreen}{2.56\%}           & \colorbox {lightgreen}{-43.23\%} \\
\bottomrule[1pt]
\end{tabular}}
\begin{tablenotes}
    \scriptsize
    \item[*] \textbf{(1)} For DeepSeek-VL2, due to its limited context window, only two methods were tested, with absolute positioning applied during assembly. \\
    \textbf{(2)} The three variants of \system correspond to different assembly strategies: using MLLMs, absolute positioning, and getting the best of the first two strategies with a dynamic verifier. \\
    \textbf{(3)} $\Delta$ represents the improvement or decline of \systemnospace’s best performance among three variances relative to the best of baselines.
\end{tablenotes}
\end{threeparttable}
\vspace{-1em}
\label{tb_results}
\end{table*}
\subsection{Evaluation Metrics}
We evaluate the generated samples using various automatic metrics in terms of \textbf{code and visual similarity} and select the following four metrics for presentation in the main text, which are more representative, relevant, and aligned with human preferences:
\begin{itemize}[leftmargin=4mm, itemsep=0.05mm] 
\item \textbf{TreeBLEU}~\cite{gui2024vision2ui}. This metric measures the structural similarity of the HTML DOM tree by calculating the recall of 1-height subtrees, relative to the reference.
\item \textbf{CLIP}~\cite{DBLP:conf/icml/RadfordKHRGASAM21}. This metric evaluates content similarity by computing the CLIP cosine similarity between the screenshot of the generated page and the original design.
\item  \textbf{Visual Score}~\cite{Si2024Design2CodeHF}\footnote{We use the implementation of visual score from the first version of their paper, which may differ from the latest version, particularly regarding the sub-indicator \textit{text color}}. A hybrid metric that calculates the match ratio of blocks and also considers block-level similarities in color, text, CLIP, and position.
\item \textbf{MAE} (\textbf{M}ean \textbf{A}bsolute \textbf{E}rror). MAE measures the average absolute pixel color value difference between images.
\end{itemize}

\subsection{Baselines}
\mypara{Backbone MLLMs.}
We use an open-source model and  two commercial models as backbone MLLMs:
\begin{itemize}[leftmargin=4mm, itemsep=0.05mm] 
\item  \textbf{DeepSeek-VL2}~\cite{Lu2024DeepSeekVLTR}.
 An open-source and advanced series of large Mixture-of-Experts Vision-Language Models which has three variants: DeepSeek-VL2-tiny, DeepSeek-VL2-small, and DeepSeek-VL2, with 1.0B, 2.8B and 4.5B activated parameters respectively.
\item \textbf{Gemini} (\texttt{gemini-v1.5-pro-latest})~\cite{DBLP:journals/corr/abs-2312-11805}.
DeepMind's Gemini specializes in seamless cross-modal reasoning, natively integrating text, code, and visual modalities through unified architecture.
\item  \textbf{ GPT-4o} (\texttt{v2024-02-01})~\cite{DBLP:journals/corr/abs-2303-08774}
 OpenAI's GPT-4o prioritizes text coherence and task generalization, leveraging massive-scale pretraining to handle complex logical chains.
 \end{itemize}.

\mypara{Comparison Methods.} We include four state-of-the-art methods and two variants of \system as baseline methods:
\begin{itemize}[leftmargin=4mm, itemsep=0.05mm] 
\item We include three baseline methods from Design2Code~\cite{Si2024Design2CodeHF}: \textbf{direct}, \textbf{text-augmented}, and \textbf{self-revision}, using exactly the same prompts and settings. 
\item \textbf{DCGen}~\cite{wanAutomaticallyGeneratingUI2024}, with \texttt{max\_depth=1} as specified in its paper.
\item \textbf{\system (APS)} and \textbf{\system (MS)}. Since our method employs two distinct strategies—Absolute Positioning Assembly (APS) and MLLM-based Assembly (MS)—for assembling blocks' code, we treat these two variants as independent baselines for comparative experiments.
\end{itemize}
Given that \cite{Si2024Design2CodeHF} and \cite{gui2024vision2ui} have demonstrated GPT-4o's superior performance over task-specified models such as WebSight-8B~\cite{Laurenccon2024UnlockingTC}, Design2Code-18B~\cite{Si2024Design2CodeHF}, and WebCoder~\cite{gui2024vision2ui}, we have excluded these models from our baseline comparisons.

\subsection{Implementation Details}
\label{sec_divider_details}
\mypara{Common Settings.}
In order to improve the reproducibility of experimental results, we set the temperature parameter of all APIs to 0 and fixed the random seeds for Python's \texttt{random}, \texttt{torch}, and \texttt{numpy} packages to 2026.
All the experiments are conducted on a Linux server equipped with 4 NVIDIA A800 80GB GPUs.

\mypara{Detailed Settings for the Dividing Algorithm.}
While we introduce the dividing algorithm in Section~\ref{sec_division}, we have omitted many trivial details to help readers focus on the key points.
We use EasyOCR\footnote{\url{https://github.com/JaidedAI/EasyOCR}} to obtain the BBoxes of all text areas and then merge adjacent BBoxes that are within 20 pixels horizontally or vertically, allowing for more accurate determination of complete text paragraphs. 
We set the grid sampling interval to 5 pixels, the minimum dividing distance ($\tau$) to 50 pixels, and the number of ignored edge points to 10. Additionally, we limit the maximum depth of recursive searches to 3. When merging blocks, we set the allowed minimum block area ($\theta$) to 300$\times$300 pixels.
These parameters ensure that the final division does not result in overly coarse or fine-grained blocks, achieving a relatively balanced visual outcome.

\section{Experimental Results and Analysis}
\subsection{Overall Performance}
\mypara{Effectiveness of \system.}
As depicted in Table~\ref{tb_results}, compared to other methods, \system with GPT-4o shows improvements of 17.65\% in TreeBLEU, 1.27\% in CLIP, 3.8\% in Visual Score, and a 37.41\% reduction in MAE on Design2Code-HARD; on \benchmark, it improves by 60\% in TreeBLEU, 2.53\% in CLIP, 2.56\% in Visual Score, and reduces MAE by 43.23\%. Similarly, with Gemini, \system shows improvements of 2.38\% in CLIP, 5.19\% in Visual Score, and a 40.09\% reduction in MAE on Design2Code-HARD, and improvements of 30\% in TreeBLEU, 2.56\% in CLIP, 2.63\% in Visual Score, and a 43.03\% reduction in MAE on \benchmark.
With DeepSeek-VL2, \system shows significant improvements in TreeBLEU (58.33\%), Visual Score (12.5\%), and MAE (reduced by 25.31\%) on Design2Code-HARD, despite a slight drop in CLIP (-4.94\%); on \benchmark, it improves TreeBLEU by 66.67\%, Visual Score by 12.5\%, and reduces MAE by 38.53\%, while CLIP decreases by -1.33\%.
These results demonstrate that \system outperforms nearly all baseline methods across all backbone MLLMs on both test benchmarks. We can conclude that \textbf{\system significantly boosts MLLMs' performance in webpage generation, especially for weaker models}.


\begin{figure}[!th]
    \centering
    \includegraphics[width=\linewidth]{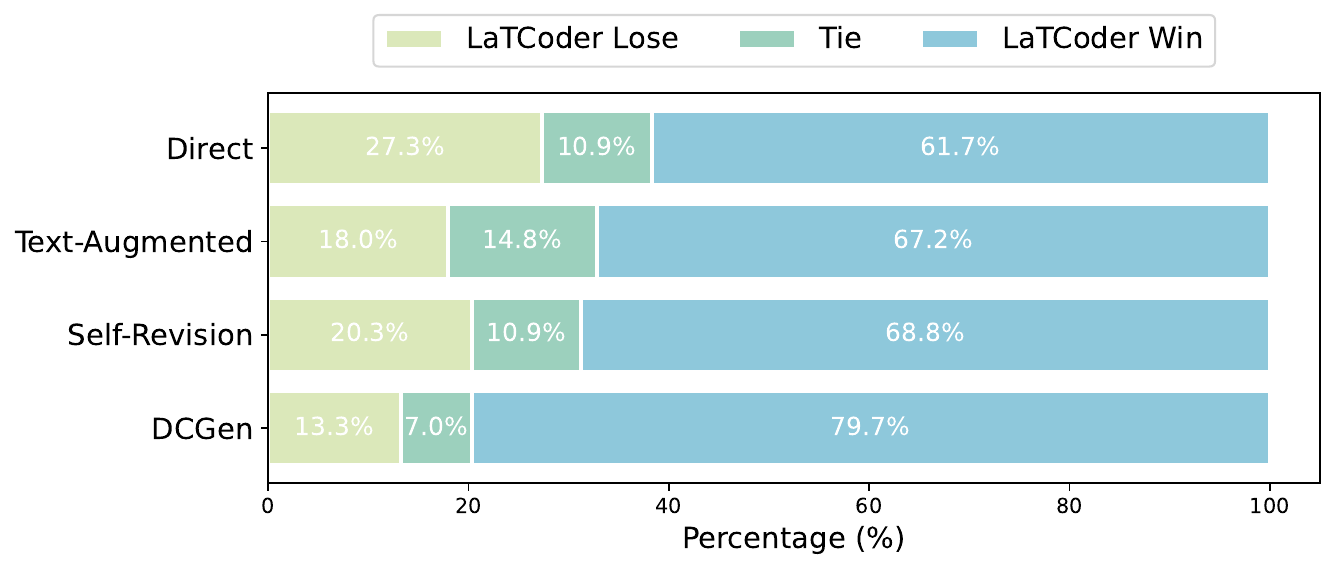}
    \caption{Pairwise human preference evaluation of baseline methods \textbf{relative to \system}, using GPT-4o as the backbone on the \benchmark dataset, with majority voting from six annotators.} 
    \vspace{-1em}
    \label{fig_preference}
\end{figure}
\mypara{\benchmark is More Challenging.}
For all three backbone models in Table~\ref{tb_results}, both \system and the baseline methods show performance drops on \benchmark, especially in TreeBLEU and MAE. 
For example, when using GPT-4o, on \benchmark compared to Design2Code-HARD, the average TreeBLEU, CLIP, and Visual Score of all methods decreased by 48.97\%, 8.03\%, and 1.24\%, respectively, while MAE increased by 9.12\%.
The results suggest that \textbf{the complexity of \benchmark is more challenging for all backbones and methods}.
We attribute the challenge of \benchmark to its more complex layouts, as shown in Table~\ref{tb_dataset_stat}, with deeper DOM trees and a greater variety and number of tags.

\subsection{Ablation Studies}
\mypara{Influence of Assembly Strategies.}
As shown in Table~\ref{tb_results}, when using Gemini and GPT-4o as backbones on both datasets, the performance differences between \system (MS) and \system (APS) in TreeBLEU, CLIP, and Visual Score are minimal, with both outperforming all baselines. This suggests that the assembly strategies have little impact on the code structure, content similarity, and block-level similarity of the generated webpage. However, \system (APS) significantly outperforms \system (MS) in MAE, likely because APS strictly preserves block positions, while MS may introduce positional changes. Using absolute positioning ensures that blocks with similar content align more precisely, reducing MAE compared to the original design.
However, we find that the results generated by MLLM-based assembly often have better overall aesthetics and smoother transitions between blocks compared to APS, which is another reason for retaining both assembly strategies. As shown in Table~\ref{tb_results}, filtering the results through a verifier further improves the generated outcomes.

\mypara{Influence of CoT-based Prompts.}
To assess the effectiveness of the CoT-based prompt for block-wise code generation, we conduct an ablation study using a simplified prompt on the \benchmark dataset, with GPT-4o as the backbone MLLM.
The simplified prompt omits any CoT-based reasoning and is as follows: \textit{you are a frontend developer, and your task is to convert a webpage screenshot into HTML and CSS code. Return format: \textquotesingle\textquotesingle\textquotesingle html code\textquotesingle\textquotesingle\textquotesingle}.
As shown in Table~\ref{tb_ablation_prompt}, the performance drops significantly when using the simplified prompt, highlighting the effectiveness of the CoT-based prompt.
\begin{table}[]
\caption{Ablation study on the CoT-based generation prompt.}
\begin{tabular}{lcccc}
\toprule[1pt]
                         & TreeBLEU & CLIP & Visual Score & MAE   \\
\midrule
CoT-based Prompt                 & 0.16      & 0.81  & 0.80          & 36.80 \\
Simplified Prompt & 0.13      & 0.76  & 0.71          & 43.17 \\
\bottomrule[1pt]
\end{tabular}
\label{tb_ablation_prompt}
\vspace{-1em}
\end{table}

\mypara{Influence of Model Scales.}
We also conduct a study on the performance of \system across different model scales on the \benchmark dataset. In this study, we use three variants of DeepSeek-VL2 as backbone MLLMs: DeepSeek-VL2-tiny, DeepSeek-VL2-small, and DeepSeek-VL2.
The results in Table~\ref{tb_model_scale} show that \system consistently achieves significant improvements, with particularly notable gains for smaller models, underscoring its general applicability.

\begin{table}[]
\caption{Performances under different model scales.}
\begin{tabular}{lcccc}
\toprule[1pt]
                                            & TreeBLEU & CLIP     & Visual Score & MAE        \\
\midrule
\multicolumn{5}{c}{\textit{\textbf{DeepSeek-VL2-tiny (3.37B, 1B activated)}}}               \\
\midrule
Direct                                      & 0.04      & 0.66      & 0.24          & 76.55      \\
\system                                    & 0.11      & 0.73      & 0.67          & 44.12      \\
$\Delta$                                          & +175\%   & +10.61\% & +179.17\%    & -42.36\%   \\
\midrule
\multicolumn{5}{c}{\textit{\textbf{DeepSeek-VL2-small (16.1B, 2.8B activated)}}}           \\
\midrule
Direct                                      & 0.08      & 0.73      & 0.61          & 61.83      \\
\system                                    & 0.12      & 0.73      & 0.68          & 49.44      \\
$\Delta$                                           & +50\%    & +0\%     & +11.48\%     & -20.04\%   \\
\midrule
\multicolumn{5}{c}{\textit{\textbf{DeepSeek-VL2 (27.5B, 4.5B activated)}}}                   \\
\midrule
Direct                                      & 0.09      & 0.75      & 0.64          & 66.91      \\
\system                                    & 0.15      & 0.74      & 0.72          & 41.13      \\
$\Delta$                                           & +66.67\% & -1.33\%  & +12.5\%      & -38.53\%   \\
\bottomrule[1pt]
\end{tabular}
\label{tb_model_scale}
\vspace{-1em}
\end{table}

\mypara{Parameters for the Dividing Algorithm.}
Since the dividing algorithm plays a crucial role and relies on several parameters, we conduct a parameter study focusing on the most critical one: the minimum area threshold for block merging ($\theta$). The experiments are conducted on the \benchmark dataset using GPT-4o as the backbone MLLM. The results in Table~\ref{tb_parameters} indicate that the threshold setting of \texttt{300*300} is close to optimal.
\begin{table}[]
\caption{Parameter study on the minimum area threshold for merging blocks in the dividing algorithm.}
\begin{tabular}{lcccc}
\toprule[1pt]
$\theta$ & TreeBLEU & CLIP & Visual Score & MAE   \\
\midrule
100*100                   & 0.14      & 0.8   & 0.75          & 37.07 \\
200*200                   & 0.14      & 0.8   & 0.76          & 36.56 \\
\textbf{300*300}               & 0.16      & 0.81  & 0.8           & 36.8  \\
400*400                   & 0.13      & 0.79  & 0.74          & 40.88 \\
500*500                   & 0.12      & 0.78  & 0.73          & 42.35 \\
\bottomrule[1pt]
\end{tabular}
\label{tb_parameters}
\vspace{-1em}
\end{table}

\subsection{Human Evaluation}
We perform a pairwise human preference evaluation of our method against all baseline methods, using GPT-4o as the backbone on the \benchmark dataset.
The annotators are asked, ``\textit{Which is the one that is more similar to the design image and of higher quality}?'' when presented with a pair of shuffled generated samples alongside the design image. 
To reduce the subjectivity of human evaluation, we apply a majority voting strategy to determine the preference for each generated sample.
The possible outcomes are classified as \texttt{win}, \texttt{tie}, or \texttt{lose}.
Results in Figure~\ref{fig_preference} show that compared to each baseline method, annotators preferred our method in at least 60\% of the cases.
Notably, when compared to DCGen, our method is favored in 79.7\% of cases.
\textbf{This comprehensive comparison provides strong evidence of the superiority of our approach}.

\begin{figure*}[!t]
    \centering
    \includegraphics[width=0.96\linewidth]{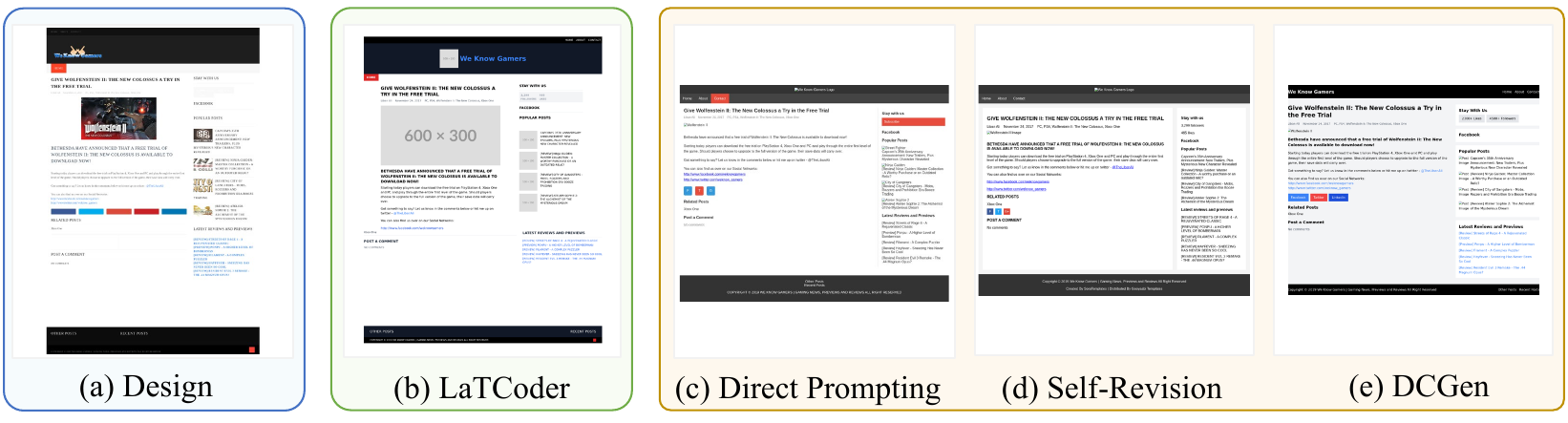}
    \caption{Case study of samples generated by \system and other baseline methods with GPT-4o as the backbone MLLM: \system significantly outperforms the others, particularly in preserving the layout of the design.} 
    \label{fig_case_study}
\end{figure*}

\subsection{Qualitative Analysis}

\mypara{Good Case Study.}
In Figure~\ref{fig_case_study}, we present a case study on samples generated by \system and other baseline methods.
It can be seen that the direct and self-revision methods exhibit very similar performance.
By contrast, DCGen, which enhances monolithic generation with natural language descriptions of local regions, even shows some performance degradation.
In comparison, \system significantly improves overall structural and detail similarity by leveraging the advantages of \textsc{LaT} and division-based generation.
Interestingly, we find that the self-revision method produces results nearly identical to those of the text-augmented method. This is likely because the self-revision method refines outputs generated by the text-augmented method, and MLLMs generally have limited capabilities for refinement.
Therefore, we omit the results from the text-augmented method for brevity.

\mypara{Error Case Analysis.}
During the experiments, we have identified two main errors that contributed to poor results:
(1) Layout misarrangement issue in the block code generation.
MLLMs sometimes incorrectly center the content of a small block at the bottom, which deviates from its intended position in the design, resulting in discrepancies between the generated output and the original design.
This highlights an inherent limitation of MLLMs in capturing precise layout information from images, even when dealing with relatively simple-styled designs. While \system significantly mitigates this issue at the overall layout level, it cannot entirely eliminate it when generating code for subregions.
(2) MLLMs' `laziness' issue.
Gemini sometimes takes shortcuts during code assembly, omitting some blocks' code and resulting in missing regions in the generated output.
We plan to explore strategies to mitigate these MLLMs' limitations in future work.

\section{Related Work}

\mypara{UI Automation.}
Early researches, constrained by limited computational resources, primarily focus on generating UI code from simple-styled design images using smaller models. For instance, pix2code~\cite{Tony2018_pix2code} leveraged LSTM and CNN architectures to produce domain-specific languages (DSLs), while Sketch2code~\cite{Alex2019_Sketch2code} explored both deep learning-based and object detection-based methods for UI prototyping from hand-drawn mockups.
With the advancement of MLLMs~\cite{DBLP:journals/corr/abs-2303-08774, DBLP:journals/corr/abs-2312-11805, TheC3}, recent studies have sought to integrate MLLMs into UI automation. Some efforts focus on curating specialized training datasets\cite{Laurenccon2024UnlockingTC, gui2024vision2ui} to enhance MLLMs' capabilities in UI generation, while others aim to establish benchmarks and evaluation metrics~\cite{Si2024Design2CodeHF, iwbench_guo_2024, DBLP:journals/corr/abs-2406-20098, xiaoInteraction2CodeHowFar2024} to systematically assess performance and drive further progress in this domain. Additionally, research~\cite{wanAutomaticallyGeneratingUI2024,zhouBridgingDesignDevelopment2024, gui2025uicopilot} has explored novel approaches to improve the visual aesthetics and interactive functionalities of generated UIs.
Despite these advancements, full UI automation remains a distant goal, particularly when dealing with real-world webpage designs that feature intricate layouts and extensive code.

\mypara{Code Intelligence.}
Neural language models have advanced code intelligence~\cite{wan2024deep}, enabling key tasks such as code summarization~\cite{wan2018improving, wang2020reinforcement}, code search~\cite{wan2019multi, DBLP:conf/wsdm/HuWD00H023}, and code generation~\cite{bi2024iterative,DBLP:conf/kbse/Sun000J0L24,ouyang2025nvagent}.
The development of code-focused LLMs has progressed through several stages. Early models like InCoder~\cite{fried2022incoder} introduced capabilities for code infilling and synthesis, enabling more flexible code generation. Further advancements were made with models such as WizardCoder~\cite{luo2023wizardcoder}, which incorporated instruction tuning to better follow complex prompts. More recent models like Qwen-Coder~\cite{hui2024qwen2} and DeepSeek-Coder~\cite{guo2024deepseek} have focused on scaling model sizes and training data, aiming to improve performance across diverse coding tasks.
While most prior work focused on general-purpose LLMs for code generation, this paper addresses a distinct problem: generating code directly from webpage designs. This task requires an understanding of UI structures, layout constraints, and code synthesis, setting it apart from conventional code generation.

\mypara{Step Reasoning in LLMs.}
Various methods have been proposed to address complex problems, aiming to mitigate hallucination issues while enhancing models' reasoning capabilities. For instance, \citet{yaoTreeThoughtsDeliberate2023} and \citet{bestaGraphThoughtsSolving2024} introduced \textit{Tree-of-Thought} (ToT) and \textit{Graph-of-Thought} (GoT), respectively, building on the foundational \textit{Chain-of-Thought} (CoT)\cite{weiChainThoughtPrompting2022} framework. 
Recently, OpenAI released a powerful commercial model, o1, which enhanced GPT's reasoning abilities by incorporating step-by-step thinking and verification processes~\cite{lightmanLetsVerifyStep2023}.
In contrast to these approaches, we propose \textsc{LaTCoder}, a method specifically tailored for the design-to-code task with layout-as-thought. \textsc{LaTCoder} preserves layout information and alleviates the burden of generating lengthy code for MLLMs.

\mypara{Layout Understanding and Generation.}
Prior work has explored layout understanding and generation across tasks like text-to-layout and document-to-layout. Text-to-layout synthesizes plausible arrangements of elements, with \cite{linLayoutPrompterAwakenDesign2023} using in-context learning for versatility and efficiency, and \cite{chaiLayoutDMTransformerbasedDiffusion2023, inoueLayoutDMDiscreteDiffusion2023} employing diffusion models~\cite{DBLP:conf/nips/HoJA20,DBLP:conf/cvpr/RombachBLEO22} for controllable generation. In document layout, \citet{Huang2022LayoutLMv3PF} pre-trained multimodal Transformers with unified text and image masking, while \citet{Appalaraju2023DocFormerv2LF} introduced DocFormerV2, trained on novel unsupervised tasks. These works, however, do not directly address \system’s goal of converting images into code.

\section{Conclusion}
In this work, we draw inspiration from the CoT reasoning in human cognition and introduce \system, a novel approach that enhances layout preservation during webpage design-to-code generation through \textit{Layout-as-Thought} (\textsc{LaT}). Specifically, we propose a simple yet effective algorithm that divides the webpage design into image blocks, instructs MLLMs to generate code for each block using CoT reasoning, and then assembles the code for all blocks using two distinct strategies with dynamic selection. 
To further evaluate the performance of MLLMs in the design-to-code task, we introduce a new and more challenging dataset, \benchmark, featuring complex layouts. We assess our method on both a public benchmark and \benchmark, with experimental results—on both automatic metrics and human evaluation—robustly demonstrating the effectiveness of our approach.

\begin{acks}
This work is supported by the Major Program (JD) of Hubei Province (Grant No. 2023BAA024).
\end{acks}

\balance
\bibliographystyle{ACM-Reference-Format}
\bibliography{ref}

\appendix

\begin{figure*}[ht]
\centering
\vspace{1em}
\begin{tcolorbox}[enhanced,attach boxed title to top center={yshift=-3mm,yshifttext=-1mm},boxrule=0.8pt, 
  colback=gray!00,colframe=black!50,colbacktitle=gray,
  title=Prompt for Block-wise Code Synthesis,
  boxed title style={colframe=gray}, left=2mm, right=2mm, top=1mm, bottom=1mm]
You are an expert Tailwind developer.

Based on the reference screenshot of a specific section of a webpage (such as the header, footer, card, etc.) provided by the user, build a single-page app using Tailwind, HTML, and JS. Please follow the detailed requirements below to ensure the generated code is accurate:

\vspace{0.3em}
\textbf{Basic Requirements:}
\begin{enumerate}
    \item \textbf{Rigid Requirements}
    \begin{itemize}
        \item You are provided with the following unmodifiable HTML framework:
        \begin{lstlisting}[language=HTML, basicstyle=\ttfamily\tiny]
        <!DOCTYPE html>
        <html lang="en">
        <head>
            <meta charset="UTF-8">
            <meta name="viewport" content="width=device-width, initial-scale=1.0">
            <script src="https://cdn.tailwindcss.com"></script>
            <link rel="stylesheet" href="https://cdnjs.cloudflare.com/ajax/libs/font-awesome/5.15.3/css/all.min.css">
        </head>
        <body>
            <!-- Your task is to fill this area -->
        </body>
        </html>
        \end{lstlisting}
        \item Your task is to generate a code block that starts with a <div> tag and ends with a </div> tag, and embed it within the <body> tag of the above-mentioned framework.
        \item Do not deliberately center the content. Arrange the elements according to their original layout and positions.
        \item The generated code should not have fixed width and height settings.
        \item Ensure that the proportions of images in the code are preserved.
        \item Both the margin and padding in the code should be set to 0.
        \item Ensure that the generated code does not conflict with outer <div> elements in terms of layout and style.
        \item The final return should be the complete HTML code, that is, including the above-mentioned framework and the code you generated and embedded into the <body> of the framework.
    \end{itemize}
    
    \item \textbf{Appearance and Layout Consistency:}
    \begin{itemize}
        \item Ensure the app looks exactly like the screenshot, including the position, hierarchy, and content of all elements.
        \item The generated HTML elements and Tailwind classes should match those in the screenshot, ensuring that text, colors, fonts, padding, margins, borders, and other styles are perfectly aligned.
    \end{itemize}
    
    \item \textbf{Content Consistency:}
    \begin{itemize}
        \item Use the exact text from the screenshot, ensuring the content of every element matches the image.
        \item For images, use placeholder images from https://placehold.co and include a detailed description in the alt text for AI-generated images.
    \end{itemize}
    
    \item \textbf{No Comments or Placeholders:}
    \begin{itemize}
        \item Do not add comments like "<!-- Add other navigation links as needed -->" or "<!-- ... other news items ... -->". Write the full, complete code for each element.
    \end{itemize}
    
    \item \textbf{Libraries to Use:}
    \begin{itemize}
        \item Use the following libraries:
        \begin{itemize}
            \item Google Fonts: Use the relevant fonts from the screenshot.
        \end{itemize}
    \end{itemize}
\end{enumerate}

\vspace{0.3em}
\textbf{Process Steps:}
\begin{enumerate}
    \item \textbf{Analyze the Section:} Based on the provided screenshot, analyze a specific section of the webpage (such as the header, footer, card, form, etc.). Break down all the elements in this section (e.g., text, images, buttons, etc.) and understand their relative positions and hierarchy.
    \item \textbf{Generate HTML Code:} Based on the analysis from Step 1, generate a complete HTML code snippet representing that specific section, ensuring all elements, positions, and styles match the screenshot.
    \item \textbf{Text Content Comparison:} Compare the generated HTML with the screenshot’s text content to ensure accuracy. If there are any discrepancies or missing content, make corrections.
    \item \textbf{Color Comparison:} Compare the text color and background color in the generated HTML with those in the screenshot. If they don't match, adjust the Tailwind classes and styles to reflect the correct colors.
    \item \textbf{Background and Other Style Comparison:} Ensure the background colors, borders, padding, margins, and other styles in the generated HTML accurately reflect the design shown in the screenshot.
    \item \textbf{Final Integration:} After reviewing and refining the previous steps, ensure that the generated HTML code is complete and perfectly matches the specific section of the screenshot.
\end{enumerate}

\vspace{0.3em}
\textbf{Code Format:} Please return the complete HTML code.
\end{tcolorbox}

\vspace{-7pt}
\caption{Prompt for block-wise code synthesis.}
\label{fig_prompt_generate}
\vspace{1em}
\end{figure*}

\end{document}